\documentclass {article}
\usepackage{wrapfig}
\usepackage{graphicx}
\usepackage{amsmath,amssymb}
\newcommand   {\half} {\frac{1} {2} }

\title {Still the ``World's Fastest Derivation of the Lorentz Transformation"}
\author
{Alan Macdonald \\
Department of Mathematics \\
Luther College, Decorah, IA,  U.S.A.\\ 
macdonal@luther.edu}
\begin {document}
\maketitle
\begin{abstract}
\noindent
Rothenstein's claim of a ``faster'' derivation of the Lorentz transformation is not correct.
\end{abstract}

\vspace{.1in}
Many years ago I published a short note ``Derivation of the Lorentz transfor\-mation''\cite{Macdonald}.
A somewhat improved version is posted on my web page under the title 
``World's fastest derivation of the Lorentz transformation'' \cite{macweb}. 
Recently Rothenstein has written 
``A faster than `World [sic] fastest derivation of the Lorentz transformation' '' \cite{Rothenstein}.

My purpose here is to present my ``fastest'' derivation and then show that Rothenstein's claim of a ``faster'' derivation is wrong.
\vspace{.05in}

Assume:

(A) The speed of light is the same in all inertial frames. (Take $c = 1$.)

(B) A clock moving with constant velocity $v$ in an inertial frame $I$ runs at a constant rate             $\gamma  = \gamma(|v|)$ with respect to the synchronized clocks of $I$ which it passes. 

Assumption (B) follows directly from the relativity principle. We do not assume that the Lorentz transformation is linear.

\begin{wrapfigure} {r}{2.25in}
\vspace{-.2in}
\centering \includegraphics[width=2.25in]{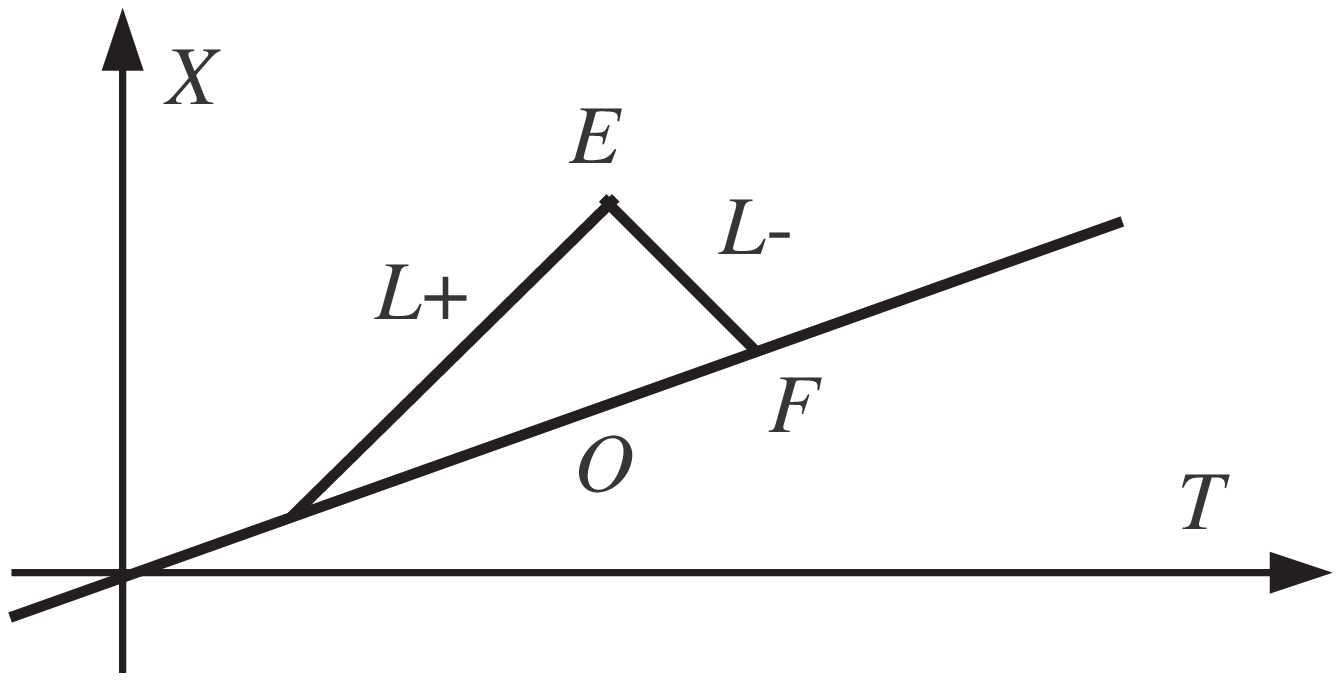} 
\vspace{-.2in}
\end{wrapfigure}

In the figure, $E$ is an arbitrary event plotted in an inertial frame 
$I$, $L+$ and $L-$ are the two light worldlines through $E$, 
and $O$ is the worldline of the spatial origin of an inertial frame $I'$ moving with velocity $v$ in $I$. 
On $O$,
\begin{align*}
	X' = 0,\ X = vT, \textrm{ and } T = \gamma T'.
\end{align*}
Thus on $O$,
\begin{align}
T + X &= \gamma(1 + v)(T' + X')	  \label{eq:tpx} \\
T - X &= \gamma(1 - v)(T' - X').	\label{eq:tmx}
\end{align}
Since $c = 1$ in $I$, an increase in $T$ along $L-$ is accompanied by an equal decrease in $X$. 
Thus $T + X$ is the same at $E$ and $F$. 
Likewise, since $c = 1$ in $I'$, $T' + X'$ is the same at $E$ and $F$. 
Thus Eq. (\ref{eq:tpx}), which is true at $F$, is also true at $E$. 
Similar reasoning using $L+$ proves Eq. (\ref{eq:tmx}) true at $E$. 
Add and subtract Eqs. (\ref{eq:tpx}) and (\ref{eq:tmx}):
\begin{align}
T &= \gamma(T' + vX')  \label{eq:t} \\
X &= \gamma(vT' + X').	\label{eq:x}
\end{align}
For $X = 0$ in Eq. (\ref{eq:x}), $X' = -vT'$; the origin of $I$ has velocity $-v$ in $I'$. 
Thus, switching $I$ and $I'$ and using (B), 
the reasoning for Eq. (\ref{eq:tpx}) also gives $T' + X' = \gamma(1 - v)(T + X)$. 
Substituting this in Eq. (\ref{eq:tpx}) gives $\gamma = (1 - v^2)^{-\half}$.
This completes the derivation of the Lorentz transformation.
\vspace{.1in}

Rothenstein writes that my ``Assumption B is a cryptic way of presenting the time dilation effect.''
This is not correct: Assumption B does not assume a specific form for the time dilation effect. 
As far as the assumption is concerned, $\gamma$ could be identically 1, meaning no time dilation. 
I \emph{derive} the time dilation factor $\gamma$ in the course of my proof of the Lorentz transformation.
On the other hand, Rothenstein assumes a specific form for the length contraction effect, his Eq. (1). 
This strong assumption is what allows his ``faster'' derivation.
In fact, a cursory comparison of the two derivations shows that his is not even faster, 
despite his strong assumption.

I pointed all this out in a reply to an email from Rothenstein, but received no reply from him.

{}

\end{document}